\newlength{\intwidth}

\documentclass[lineno]{jfm}

\usepackage{amssymb}

\usepackage{array}
\usepackage{amsmath}
\usepackage{latexsym}
\usepackage{bezier}
\usepackage{psfrag,graphicx}
\usepackage{bm}
\usepackage{float}
\usepackage{multirow}
\usepackage{mathrsfs}
\usepackage{color}
\usepackage{enumerate}
\usepackage{natbib}
\usepackage{nicefrac}
\usepackage[colorlinks, citecolor=blue]{hyperref}
\usepackage{overpic}
\usepackage{rotating}
\usepackage{verbatim}

\begin{document}

\title[Finite amplitude waves in G\"ortler vortices]{Beyond secondary instability:  on the emergence of finite-amplitude waves in G\"ortler vortices 
}

\author{
  Runjie Song\aff{1}
  \and Kengo Deguchi\aff{1}
}
 
 \affiliation{
   \aff{1}School of Mathematics, Monash University, VIC 3800, Australia
}

\maketitle

\begin{abstract}
G\"ortler vortices developing over a concave wall \textcolor{black}{support} rapidly oscillating wavelike disturbances through secondary instabilities. Although experiments indicate that the finite-amplitude evolution of these waves acts as a precursor to turbulence transition, accurate and efficient prediction has remained out of reach. We overcome this limitation by using the Parabolised Coherent Structures (PCS) method of \cite{Song_Deguchi_2025}, which 
incorporates the nonlinear vortex-wave interaction into a standard spatial-marching approach.
\textcolor{black}{Our computations successfully reproduce the wave amplitude and displacement thickness observed in the widely known experiments of \cite{Swearingen_Blackwelder_1987}.}
\end{abstract}

\section{Introduction}\label{sec:introduction}

G\"ortler vortices are streamwise-alighned vortical structures that arise within the boundary layer over a concave wall \citep{Gortler_1940}. Their significance, extensively reviewed for example by \cite{Floryan_1991}, \cite{Saric_1994}, \textcolor{black}{and \cite{Xu_2024},} continues to draw attention due to their critical role for the design of high-speed vehicles and jet engines
\citep{Li_et_al_2022,Es-Sahli_2022,Xu_Ricco_Marensi_2024,Zhang_Hao_Uy_2025}. Pioneering experiments by \cite{Swearingen_Blackwelder_1987}, hereafter referred to as SB87, demonstrated that the downstream transition process involves the gradual development of G\"ortler vortices, followed by the emergence of short-scale wavy fluctuations that eventually break down into turbulence. 
Although the initial stages of the transition process are relatively well understood, a complete understanding has not yet been achieved.

\begin{figure}
\centering
\begin{overpic}[width=0.8 \textwidth]{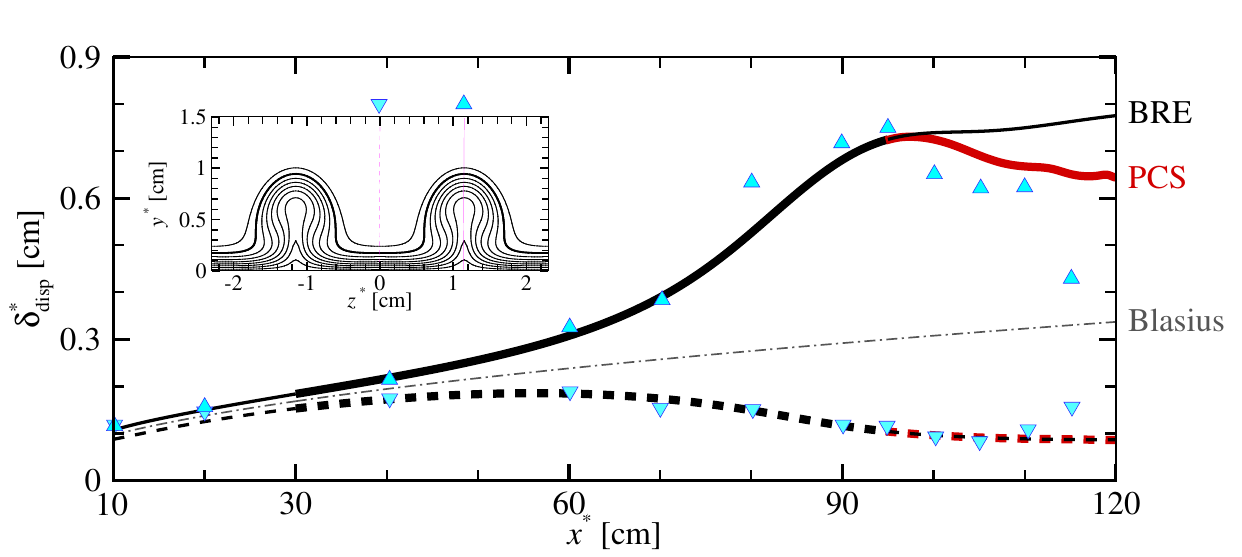}
\end{overpic}
\caption{Displacement boundary layer thickness $\delta^*_{\text{disp}}$ measured at the spanwise locations corresponding to the peaks (up triangle, solid lines) and valleys (down triangle, dashed lines) of the G\"ortler vortices. The symbols are the experimental results taken from figure 9 of SB87, while the lines show our computational results. The inset shows the streamwise velocity from the BRE computation at $x^*=90$[cm] (in the same format as figure~\ref{fig:fig4_new}). The magenta lines and symbols indicate the positions of the peaks and valleys of the mushroom-shaped vortices.}
\label{fig:fig1}
\end{figure}
Asymptotic analysis has long been an invaluable tool for examining the initial stages of transition. A milestone was \cite{Hall_1983}, who pointed out that, \textcolor{black}{unless the local G\"ortler number is sufficiently large,}
linear growth of G\"ortler vortices
could not be determined solely from the eigenvalue analysis of the local flow field. \textcolor{black}{In such cases,}  numerical integration of a parabolic leading order problem along the streamwise direction is necessary. \cite{Hall_1988} solved the nonlinear version of the spatial-marching problem, which is now known as the boundary region equations (BRE). As shown by the black line in figure~\ref{fig:fig1}, the BRE predictions agree remarkably well with the experiments of SB87 (symbols) up to a certain streamwise location.

In the numerical boundary layer community,
the discrepancy in figure~\ref{fig:fig1} beginning at $x^*=$95 [cm] is believed to be caused, at least in part, by secondary instability. 
\cite{Hall_Horseman_1991} showed that when nonlinear G\"ortler vortices emerge, the streamwise velocity exhibits pronounced streaky modulations, which in turn trigger inviscid instabilities. In subsequent studies, spatial marching computations of G\"ortler vortices, combined with their local linear stability analysis, have become a standard framework \citep{Yu_Liu_1994,Li_Malik_1995,Girgis_Liu_2006,Ren_Fu_2015,Xu_Zhang_Wu_2017}. However, stability analysis applies only to very small wave amplitudes, and no quantitative study has examined the nonlinear effects of finite-amplitude waves on the flow.

In this paper, we addresses this challenge using the parabolised coherent structures (PCS) method recently proposed by \cite{Song_Deguchi_2025}. 
To incorporate the feedback from fast-scale finite-amplitude waves into the slowly developing vortices within a spatial marching scheme, the PCS couples the BRE with the computation of \textit{exact coherent structures}. This approach is motivated by recent studies of parallel shear flows \citep{wang2007,HALL_SHERWIN_2010,DEGUCHI_HALL_2014a}, which revealed that simple nonlinear Navier–Stokes solutions (i.e. exact coherent structures) embody the vortex–wave interaction (VWI, \cite{hall1991}), also known as the self-sustaining process of coherent structures \citep{hamilton1995,waleffe1997}. Our aim is to demonstrate that the detailed understanding of coherent structures accumulated within the asymptotic and dynamical systems theory community, is indeed useful for explaining experimental results by SB87, even beyond the point at which secondary instabilities arise.

The paper is organised as follows. \S \ref{sec:2} presents the formulation of the PCS, along with its justification via asymptotic analysis and the details of the computational setup. \S \ref{sec:3} reports the computational results and compares them with the experiment of SB87. 
\textcolor{black}{We mainly use  experimental data because  incompressible direct numerical simulation (DNS) studies that record wave amplitudes are rare; one such exception is \cite{Souza_2017}.
Finally, \S \ref{sec:4} briefly \textcolor{black}{discusses the implications of our computations. 
}
}

\section{Formulation of the problem}\label{sec:2}

Consider a boundary layer developing over a concave wall with a dimensional radius of curvature $a^*$. The incompressible Navier-Stokes equations in cylindrical coordinates $(r^*,\theta,z^*)$ serve as the starting point of our analysis. In boundary layer flows, the Reynolds number is typically defined as $Re=\frac{U_{\infty}^*L^*}{\nu^*}$, where $U_{\infty}^*$ is the freestrem velocity, $L^*$ is a characteristic streamwise length scale, and $\nu^*$ is the kinematic viscosity of the fluid. The flow field is assumed to be periodic in $z^*$ with a period of $\lambda^*$.

By using the characteristic boundary layer 
thickness $\delta^*=Re^{-1/2}L^*$ as the length scale, $U_{\infty}^*$ as the velocity scale, and the fluid density $\rho^*$ as the density scale, one obtains the Navier-Stokes equations for the velocity $\mathbf{u}=u_r\mathbf{e}_r+u_{\varphi}\mathbf{e}_{\varphi}+u_z\mathbf{e}_z$ and pressure $p$ fields. In this well-known non-dimensional formulation in $(r,\varphi,z)$, where $r=r^*/\delta^*$ and $z=z^*/\delta^*$, the viscous terms appear multiplied by $1/R_{\delta}$, where $R_{\delta}$ is the Reynolds number based on the boundary layer thickness:
\begin{eqnarray}
R_{\delta}=\frac{U_{\infty}^*\delta^*}{\nu^*}=Re^{1/2}.
\end{eqnarray}
Further setting $[u,v,w]=[u_{\varphi},u_r,u_z]$, 
$y=r-a$, and $x=a\varphi$ with $a=a^*/\delta^*$, the governing equations are recast in a suitable coordinate system, that forms the basis for deriving the reduced equations for the PCS. \textcolor{black}{For convenience, we introduce $x^*=\delta^*x$, which represents the dimensional distance from the leading edge.}

The VWI theory by \cite{hall1991} and \cite{HALL_SHERWIN_2010} provides the most rational way to identify the dominant terms in the governing equations when $Re\gg 1$. However, numerical computation of the resulting leading–order system is highly challenging. 
As we shall see in \S \ref{sec:2.2}, the PCS method therefore employs a reduced system that lies between the full equations and the VWI system. \textcolor{black}{In other words, the PCS retains some higher-order terms, which is advantageous both for simplifying numerical computations and for improving accuracy of the approximation (see \cite{Song_Deguchi_2025}).
}

\subsection{Vortex-wave interaction}\label{sec:2.1}

In the VWI theory, the velocity and pressure fields are decomposed as
\begin{eqnarray}
[\mathbf{u},p]=[\overline{\mathbf{u}},\overline{p}](X,y,z)+[\tilde{\mathbf{u}}, \color{black}{ \tilde{p}]}(X,\theta,y,z),\label{veldecomp}
\end{eqnarray}
using the slow variable 
$X=R_{\delta}^{-1}x=x^*/L^*$ and 
the fast phase variable
\begin{eqnarray}\label{phaseeq2}
\theta=R_{\delta}\int^X_0 \alpha(s)ds- \Omega t.
\end{eqnarray}
Here, $\alpha(X)$ is the local wavenumber, and $\Omega$ is the frequency; both are assumed to be real-valued. The overline denotes the steady vortex part, while the tilde denotes the time-dependent wave part; their asymptotic expansions are given by
\begin{subequations}\label{vortexVWIexp}
\begin{eqnarray}
[\overline{u},\overline{v},\overline{w},\overline{p}]&=&[U,R_{\delta}^{-1}V,R_{\delta}^{-1}W,R_{\delta}^{-2}P](X,y,z)+\cdots,\\
~[\tilde{\mathbf{u}},\tilde{p}]&=&\epsilon^{1/2} R_{\delta}^{-1}e^{i\theta}[\hat{\mathbf{u}},\hat{p}](X,y,z)+\text{c.c.}+\cdots,
\end{eqnarray}
\end{subequations}
where c.c. stands for complex conjugate, and $\epsilon=R_{\delta}^{-1/3}$ denotes the critical layer thickness. 
In the self-sustaining process, $U$, and $[V,W]$ are referred to as the streak and roll components, respectively. \textcolor{black}{As shown by \cite{hall1991} and \cite{HALL_SHERWIN_2010}, a wave amplitude of $O(R_{\delta}^{-7/6})$ is required to drive $O(1)$ streaks.}

By substituting (\ref{vortexVWIexp}) into the governing equations, defining the G\"ortler number as $G=2R_{\delta}^2/a$, and retaining only the leading-order terms, we obtain
\begin{subequations}\label{BREBRE}
\begin{eqnarray}
[U\partial_X +V\partial_y +W\partial_z-\partial_y^2-\partial_z^2]
\left[ \begin{array}{c} U\\ V\\ W \end{array} \right]
+ \left[ \begin{array}{c} 0\\ \partial_y P+G U^2/2 \\ \partial_z P  \end{array} \right] =\mathbf{F},~~~~~~\label{BRE}\\
\partial_XU+\partial_yV+\partial_zW=0,~~~~~~~
\end{eqnarray}
\end{subequations}
from the mean part and 
\begin{subequations}\label{Rayleigh}
\begin{eqnarray}
i\alpha(U-c)\hat{u}+\hat{v}\partial_y U+\hat{w}\partial_z U=-i\alpha \hat{p},\qquad
i\alpha(U-c)\hat{v}=-\partial_y \hat{p},\\
i\alpha(U-c)\hat{w}=-\partial_z \hat{p},\qquad
i\alpha \hat{u}+\partial_y\hat{v}+\partial_z\hat{w}=0.
\end{eqnarray}
\end{subequations}
from the fluctuation part. The right-hand side of (\ref{BREBRE}) is the Reynolds stress produced by the wave and we retain the higher-order terms for later use; without this term, (\ref{BREBRE}) reduces to the BRE used in \cite{Hall_1988}.
While (\ref{Rayleigh}) represents the inviscid secondary instability equations used by \cite{Hall_Horseman_1991}, with $c(X)=\Omega/\alpha(X)$ being the local phase speed. 

Careful analysis shows that the feedback from the wave to the roll-streak field occurs solely within the critical layer of thickness $\epsilon$, which regularises the singularity of (\ref{Rayleigh}) at $U=c$ \textcolor{black}{by reintroducing the viscous effect. 
The method of Frobenius expansion shows that the eigenfunction of the inviscid wave problem becomes singular (see e.g. \cite{Deguchi_2019}).
The regularised wave within the critical layer thus acquires a larger size, and the magnitude of $\mathbf{F}$ increases accordingly. }
After a long algebra, it can be shown that the Reynolds stress term affects $V,W,P$ through jump conditions across the critical layer (see (2.15) and (B 16) of \cite{Song_Deguchi_2025}, for example).
\textcolor{black}{The} 
$\epsilon^{1/2}$ factor in (\ref{vortexVWIexp}) arises from the stress balance within the critical layer.

The jump conditions, together with the BRE and (\ref{Rayleigh}), form a closed system determining the leading-order approximation of the solution. However, numerical computation of this problem requires handling the singular behaviour of the solutions at the critical layer, whose location is not known a priori. For this reason, even more than 30 years after the leading-order problem was derived for boundary-layer flows by \cite{hall1991}, no solution has yet been obtained.

\subsection{Parabolised coherent structures method}\label{sec:2.2}

In the PCS approach, the jump conditions in the VWI are replaced by the  Reynolds stress
\begin{eqnarray}
\mathbf{F}=-[R_{\delta}\overline{(\tilde{\mathbf{u}}\cdot \tilde{\nabla})\tilde{u}},R_{\delta}^2\overline{(\tilde{\mathbf{u}}\cdot \tilde{\nabla})\tilde{v}},R_{\delta}^2\overline{(\tilde{\mathbf{u}}\cdot \tilde{\nabla})\tilde{w}}],\label{Rstress}
\end{eqnarray}
and (\ref{Rayleigh}) is \textcolor{black}{replaced by}
\begin{subequations}\label{waveeqq}
\begin{eqnarray}
(\overline{\mathbf{u}}\cdot \tilde{\nabla}-c\alpha  \partial_{\theta})\tilde{\mathbf{u}}+
(\tilde{\mathbf{u}}\cdot \tilde{\nabla})\overline{\mathbf{u}}
+(\tilde{\mathbf{u}}\cdot \tilde{\nabla})\tilde{\mathbf{u}}-\overline{(\tilde{\mathbf{u}}\cdot \tilde{\nabla})\tilde{\mathbf{u}}}
=-\tilde{\nabla} \tilde{p}+R_{\delta}^{-1}\tilde{\nabla}^2 \tilde{\mathbf{u}},~~~\label{waveeq}\\
\tilde{\nabla}\cdot \tilde{\mathbf{u}}=0,~~~~~~~\label{wave2}
\end{eqnarray}
\end{subequations}
where $\tilde{\nabla}=[\alpha \partial_{\theta},\partial_y,\partial_z]$.
The PCS approach solves (\ref{BREBRE}), (\ref{Rstress}), and (\ref{waveeqq}) with 
$[\overline{u},\overline{v},\overline{w},\overline{p}]=[U,R_{\delta}^{-1}V,R_{\delta}^{-1}W,R_{\delta}^{-2}P]$.
\textcolor{black}{The PCS system can be derived directly from the Navier-Stokes equations by making the following assumptions:
(i) The effect of wall curvature on the flow field enters solely through the so-called G\"ortler term; 
(ii) the mean part of the governing equations can be parabolised; and
(iii) derivatives with respect to the slow variable $X$ can be neglected when acting on the fluctuation components.}

The PCS system possesses some remarkable properties. First, its asymptotic analysis yields the same leading-order VWI system. Thus, the PCS computes approximate solutions when $Re\gg 1$. A detailed discussion on the relationship between the VWI theory and the PCS method can be found in \cite{Song_Deguchi_2025}. \textcolor{black}{As discussed in that paper, assumptions (i)--(iii) can be justified through asymptotic analysis.}
\textcolor{black}{Second, the PCS formulation allows spatial marching (i.e. integration in $X$), similar to the BRE. Hence, the numerical computation is significantly more efficient than obtaining a statistically steady state via a DNS.
Third,} if the system is independent of $X$, as is the case in parallel flows, travelling-wave solutions of the Navier-Stokes equations satisfy it exactly. This allows the self-sustaining process of exact coherent structures to be naturally incorporated into spatial marching. 
Fourth, under the constant freestream speed considered here, the system admits the Blasius boundary-layer solution. The PCS, therefore, do not require artificial forcing to maintain the base flow, unlike local periodic box computations of spatially developing flows \textcolor{black}{(e.g. \cite{Kozul_Chung_Monty_2016,Ruan_2021}).}

Using an implicit finite-difference scheme in the $X$-direction, updating the fields is equivalent to solving a travelling-wave problem in a parallel flow, with contributions from the previous step treated as a forcing term. 
Thus \cite{Song_Deguchi_2025} builds on a code 
that computes exact coherent structures by finding the root of a discretised system using the Newton-Raphson method. The descretisation uses Fourier-Galerkin in the $\theta$ and $z$ directions, and Chebyshev collocation in the $y$ direction. 

Besides the modifications due to the finite-difference in the $X$-direction, a few more adjustments are necessary. In the numerical computations, the $y$-domain must be truncated at a finite value $H$.  
For accurate computation, $H$ must be chosen large enough while ensuring sufficient resolution within the boundary layer. To satisfy both requirements, the Chebyshev expansion is first performed in $Y\in [-1,1]$ and then mapped to $y\in [0,H]$ via $y=\frac{HB(1+Y)}{(1-Y)H+2BY}$. By choosing the constant $B$ to be sufficiently smaller than $H/2$, the Chebyshev collocation points can be clustered near the wall at $y=0$. 
No-slip boundary conditions are imposed at $y=0,H$ on the perturbation to the Blasius boundary layer.

Also, in the Newton method for exact coherent structures, $\alpha$ is prescribed and $c$ is obtained as part of solution. In the PCS, however, the product $\alpha c$ must equal to the prescribed constant $\Omega$, which provides a natural update condition for $\alpha(X)$. 
It should also be noted that, unlike in the parabolised stability equations (PSE) approach, $\alpha$ is purely real. 
A brief comment on the difference between the BRE and the PSE can be found in \cite{Xu_Wu_2021} and \cite{Song_Deguchi_2025}.

\subsection{Computational set up}

To simulate the experimental conditions of SB87, we choose $U_{\infty}^*=5$ [m/s], $a^*=3.2$ [m], $\lambda^*=2.3$ [cm] and $\nu^*=1.6\times 10^{-5}$ [m$^2$/s]. The kinematic viscosity is estimated for dry air around 288K. Using $L^*=10$ [cm], we obtain $Re=31250$ and $G=11.0485$. The map parameters are $(H,B)=(900,40)$, which place the upper boundary of the domain at $y^*=50.9$ [cm]. Most of the computations are performed using up to 150th Chebyshev modes in $y$ and 24th Fourier harmonics in $z$. Resolution was verified by increasing these numbers to 210 and 28, respectively. Following the VWI theory, only a single Fourier mode is retained in $\theta$. Including up to the second harmonic did not quantitatively change the results, except for a slight effect on the stability of the marching procedure.

We follow the approach of \cite{Hall_Horseman_1991} for generating G\"ortler vortices. 
First, the linearised BRE is solved from $x^*=10$ [cm] to 30 [cm] (thin solid and dashed lines in figure \ref{fig:fig1}) using a simple analytic initial perturbation of the form $A_0 y^6\exp(-y^2/2X)$ added to the Blasius profile. The nonlinear BRE is then solved from $x^*=30$ [cm] onward (thick lines). The amplitude $A_0$ in the initial condition is adjusted so that the numerical results at $x^*=40$ [cm] matches the experimental data in figure \ref{fig:fig1}. 
\textcolor{black}{The choice of the analytic initial profile is somewhat arbitrary, as it is unclear which conditions are actually realised in the SB87 experiments.} 
Following \cite{Hall_1983}, we checked slightly different analytical initial conditions, but our conclusions remained unchanged. 

The secondary instability can be analysed by substituting the BRE solution for $\overline{\mathbf{u}}$ in (\ref{waveeqq}) and neglecting the quadratic terms in $\tilde{\mathbf{u}}$. \textcolor{black}{The simplest way to obtain a finite-amplitude wave solution in the PCS system is to introduce a small external forcing near the streamwise location at which the secondary instability analysis indicate neutrality \citep{Song_Deguchi_2025}.
The resulting solution is essentially unaffected by the imposed forcing except in the immediate vicinity of the linear critical point, as shown in Appendix \ref{sec:AppA} (that is, the structure and amplitude of the forcing are not physically important).}

\section{Results}\label{sec:3}

\begin{figure}
\centering
\begin{overpic}[width=0.9 \textwidth]{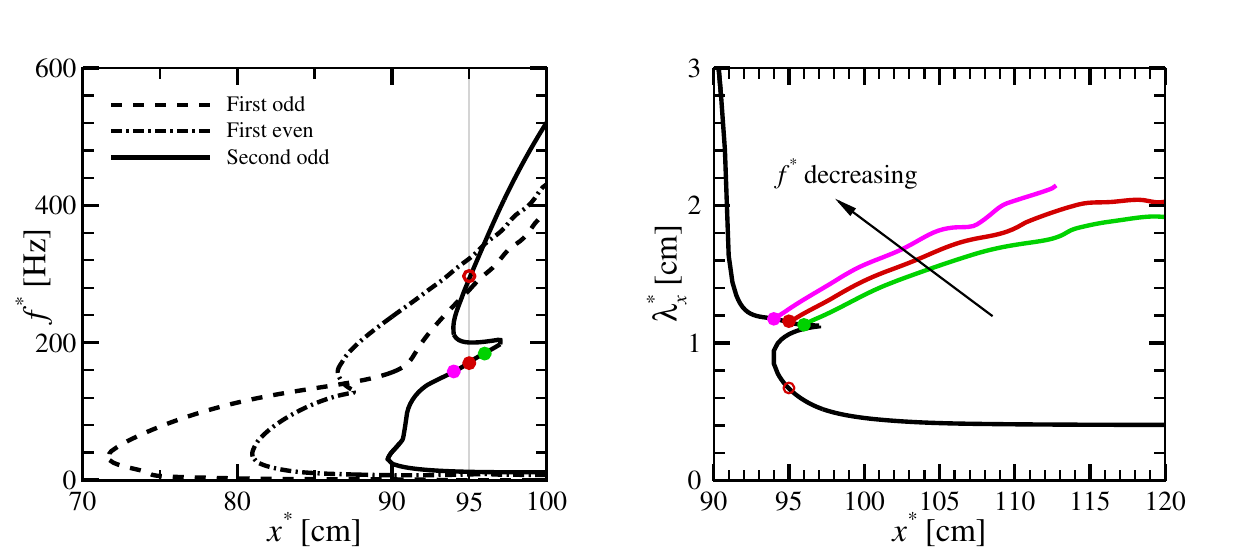}
\put(-3,39.3){(a)}
\put(47,39.3){(b)}
\end{overpic}
\caption{
(a) Neutral curve in the $x^*$--$f^*$ plane resulting from the secondary instability analysis. 
The filled circles are the linear critical points used in the PCS computations in panel (b). The open circle corresponds to the analysis in figure \ref{fig:high_low_different_R_combine}-(b).
(b) Local wavelength of the finite-amplitude wave obtained using the PCS method (the magenta, red, and green lines correspond to $f^* = 157$, 170, and 185 [Hz], respectively). The black line shows the secondary instability analysis results for the second odd mode. 
}
\label{fig:fig2}
\end{figure}

In figure \ref{fig:fig1}, for $x^*< 95$ [cm], the fluctuation field is zero, so the flow corresponds to the G\"ortler vortex captured by the BRE. Both the displacement thickness and the flow fields at $x^*=60$, 80, and 90 [cm] are in excellent agreement with the results presented in SB87, consistent with the observations of \cite{Hall_Horseman_1991} and 
\cite{Xu_Zhang_Wu_2017}. 

To obtain the PCS result (red lines in figure \ref{fig:fig1}), the secondary instability of the G\"ortler vortex must first be analysed.
As reported in the previous studies, the  instability exhibits three modes: the first odd, the first even, and the second odd modes (see figures 19 and 21 in \cite{Xu_Zhang_Wu_2017}, for example). 
Figure \ref{fig:fig2}-(a) shows the neutral curve obtained from our eigenvalue analysis. 
At each $x^*$, we computed the values of $\alpha$ and $c$ that yield neutrality \textcolor{black}{(hereafter referred to as the linear critical point)}, and plotted the corresponding frequency $f^*=\frac{ U^*_{\infty}}{2\pi \delta^*}c\alpha$ [Hz]. 
The first odd, first even, and second odd modes instability appear around $x^* = 70$, $80$, and $90$ [cm], respectively.
At $x^* = 95$ [cm], several linear critical points exist. One of these, corresponding to the second odd mode and marked by the red circle, has a frequency comparable to the 130 [Hz] observed by SB87 at $x^* = 100$ [cm]. Hence, the vicinity of this point provides a suitable starting location for the PCS spatial marching including the finite-amplitude waves.

The magenta, red and green curves in figure \ref{fig:fig2}-(b) show the PCS results obtained using the frequencies $f^* = 157$, 170, and 185 [Hz], respectively.
The corresponding \textcolor{black}{linear critical} points are approximately at $x^* = 94$, 95, and 96 [cm]. 
Recall that in the PCS calculations, we must fix $\Omega = \frac{2\pi \delta^*}{ U^*_{\infty}}f^*$, and this condition yields the local wavenumber $\alpha$ at each streamwise location. 
The corresponding local wavelengths \textcolor{black}{$\lambda_x^* = 2\pi \delta^*/\alpha$} shown in figure \ref{fig:fig2}-(b) are comparable to the 2–2.5 [cm] reported in SB87. 
All subsequent calculations in this section and the results shown in figure \ref{fig:fig1} use the PCS solution beginning at $x^* \approx 95$ [cm] (the red line, $f^*=170$ [Hz]), unless otherwise stated.  Computations starting too far upstream terminate before reaching $x^* = 120$ [cm], as indicated by the magenta line. Conversely, if the starting point is chosen too far downstream, the \textcolor{black}{linear critical point} corresponding to the relevant frequency disappears (see figure \ref{fig:fig2}-(a)).

\begin{figure}
\centering
\begin{overpic}[width=0.9 \textwidth]{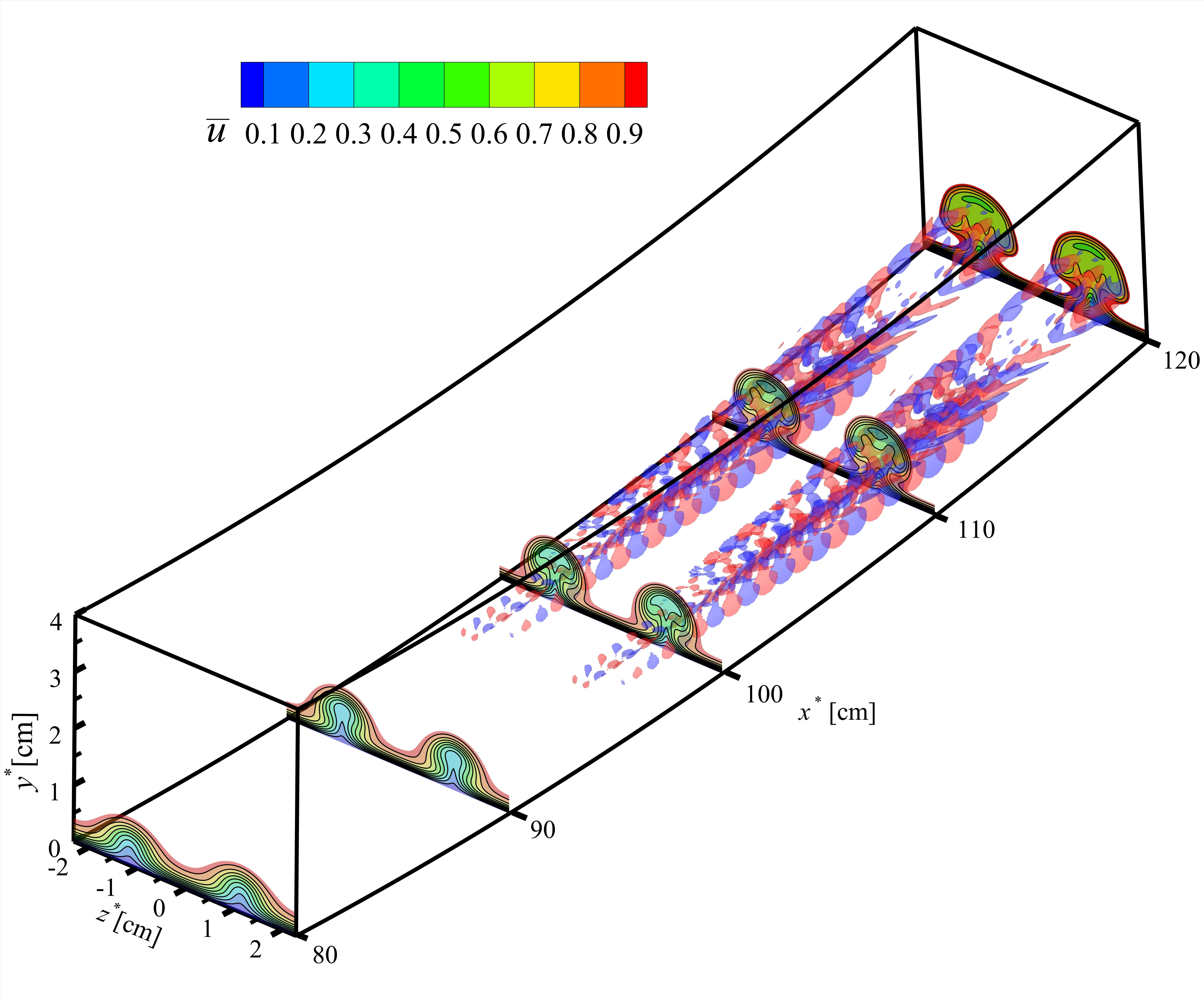}
\end{overpic}
\caption{A snapshot of the flow field computed using the PCS. The colourmap at the selected streamwise positions shows the steady streak field $\overline{u}$.
Red/blue isosurfaces are 20\% maximum/minimum of the streamwise vorticity of the wave component, $\partial_y\tilde{w}-\partial_z \tilde{v}$. 
}
\label{fig:fig3}
\end{figure}

Figure \ref{fig:fig3} illustrates the flow field obtained from the PCS computation.
Recall that in the decomposition (\ref{veldecomp}), the roll-streak part $\overline{\mathbf{u}}$ is stationary, while the \textcolor{black}{wave} field $\tilde{\mathbf{u}}$ is time-periodic with period $2\pi/\Omega$. The colourmap of $\overline{u}$ depicts the development of the G\"ortler vortices, which nonlinearly interact with the finite-amplitude waves represented by the isosurfaces. 

\begin{figure}
\centering
\begin{overpic}[width=0.95 \textwidth]{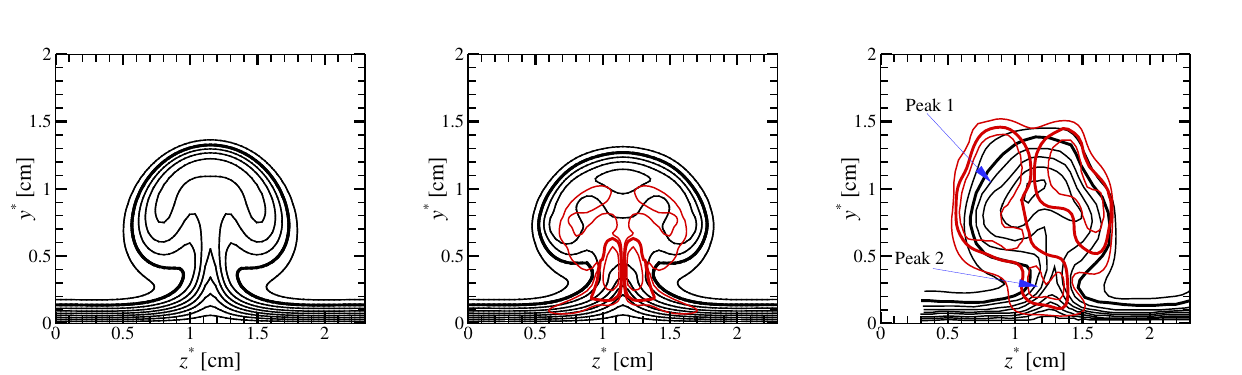}
\put(-3,25){(a)}
\put(31,25){(b)}
\put(65,25){(c)}
\end{overpic}
\caption{
The flow field at $x^*=100$ [cm]. The black lines show contours of $\overline{u}$ at $0.1,0.2,\dots,0.9$, with the thick line indicating 0.8. The red lines are contours of $\tilde{u}_{\text{rms}}=0.01,0.02,$ and 0.03, with the thick line highlighting 0.02. (a) BRE, (b) PCS, (c) experimental results from figures 11 and 16 of SB87.
}
\label{fig:fig4_new}
\end{figure}

The black contours in figures \ref{fig:fig4_new}-(a) and (b) compare the BRE and PCS results for $\overline{u}$ at $x = 100$ [cm]. The interaction between the vortex and the wave reduces the displacement thickness at the peak of the mushroom vortices, as seen in figure \ref{fig:fig1}. \textcolor{black}{Figure \ref{fig:fig4_new}-(a)} presents the experimental results from SB87. The red contours in this figure, representing the root-mean-square velocity of the wave field
\begin{eqnarray}\label{equrms}
\tilde{u}_{\text{rms}}(X,y,z)=\sqrt{\frac{1}{2\pi}\int^{2\pi}_0\tilde{u}^2 d\theta},
\end{eqnarray}
exhibit two peaks indicated by the arrows. The structures of peaks 1 and 2 resemble the first and second odd modes of the secondary instability, respectively \citep{Xu_Zhang_Wu_2017}. As expected, $\tilde{u}_{\text{rms}}$ of the PCS result, continued from the second odd modes and shown in panel (b), closely resembles peak 2.

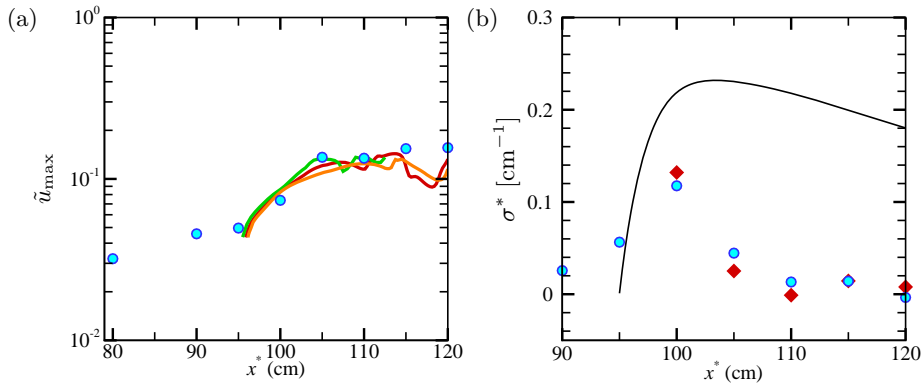
\begin{figure}
\centering
\begin{overpic}[width=0.9 \textwidth]{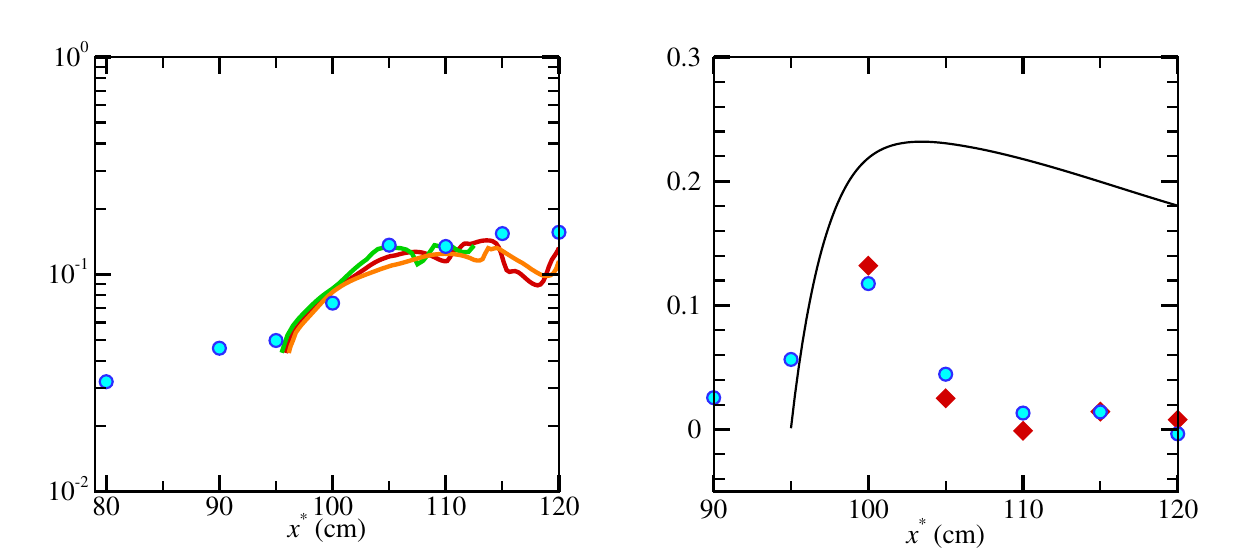}
\put(0,20){\begin{turn}{90}{$\tilde{u}_{\text{max}}$}\end{turn}}
\put(50,18){\begin{turn}{90}{$\sigma^{*}$ [cm$^{-1}$]}\end{turn}}
\put(-3,39.3){(a)}
\put(47,39.3){(b)}
\end{overpic}
\caption{
Downstream growth of the wave amplitude. (a) Amplitude of the \textcolor{black}{wave} field.
Circles denote the experimental results taken from figure 17 in SB87. Lines are the PCS results shown in figure \ref{fig:fig2}-(b).
(b) Growth rate $\sigma^*=\sigma/L^*$, where $\sigma(X)=\frac{1}{\tilde{u}_{\text{max}}}\frac{d \tilde{u}_{\text{max}}}{dX}$. For a fair comparison, finite-difference approximations are applied to both the experimental (circles) and PCS (diamonds) results. The line shows the growth rate \textcolor{black}{of the second odd mode} predicted by the secondary instability analysis. Both computational results \textcolor{black}{are for} $f^*=170$ [Hz]. 
}
\label{fig:fig5}
\end{figure}

SB87 noted that peak 1 emerges first, followed by the development of peak 2 for $x^*\geq 90$ [cm], consistent with figure \ref{fig:fig2}-(a). Although peak 2 was less dominant at $x^*=100$ [cm] in figure \ref{fig:fig4_new}-(c), it ultimately attained the highest amplitude in the experiments. The downstream growth of this peak can be quantified by the wave amplitude $\tilde{u}_{\text{max}}(X)=\max_{y,z}\tilde{u}_{\text{rms}}$.
As figure \ref{fig:fig5}-(a) indicates, the amplitude obtained by PCS agrees very well with the experimental results of SB87.

The experiments of SB87 strongly suggest that both of the two odd modes participate in the nonlinear interaction simultaneously. While it is possible to incorporate such multiple finite-amplitude waves in the PCS, this lies beyond the scope of the present work.

In  \cite{Yu_Liu_1994}, the seemingly linear growth around $x^* = 100$ [cm] in figure \ref{fig:fig5}-(a) has been attributed to secondary instability. However, as shown in figure \ref{fig:fig5}-(b), the dimensional growth rate  predicted by the secondary instability analysis ($\sigma^* = -\Im(\alpha)/\delta^*$, indicated by the line) tends to be significantly larger than that observed experimentally; a similar conclusion was reached by \cite{Xu_Zhang_Wu_2017}.
On the other hand, the PCS results are much closer to the experimental observations (red diamonds). The corresponding values of $\sigma^*$ are computed from the spatial evolution of the wave amplitude $\tilde{u}_{\text{max}}(X)$ defined earlier; see the figure caption also. \textcolor{black}{The physical roles of PCS and secondary instabilities in Navier-Stokes computations will be discussed in section \ref{sec:4}.}

\begin{figure}
\centering
\begin{overpic}[width=0.9 \textwidth]{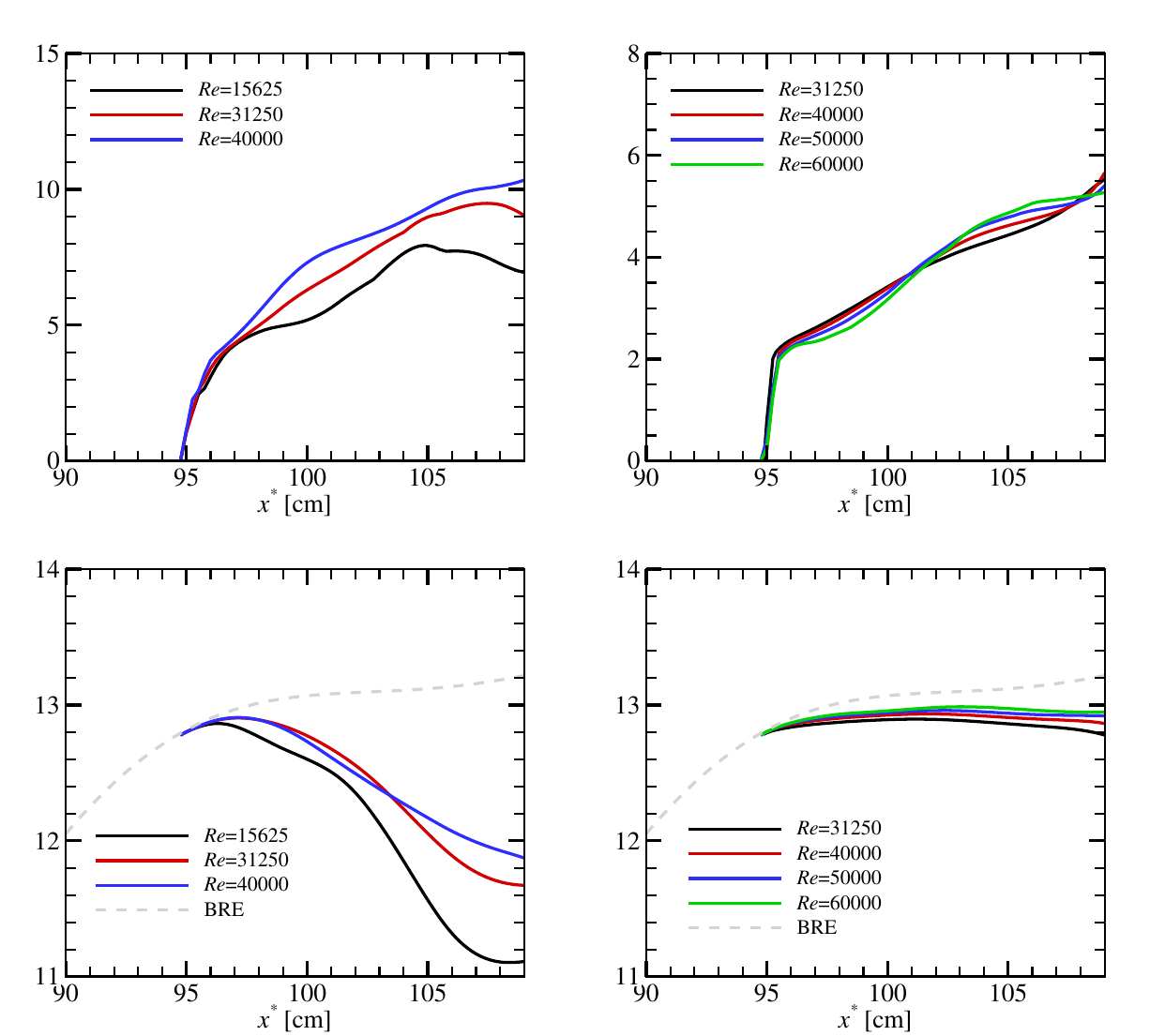}
\put(-3,84){(a)}
\put(47,84){(b)}
\put(-3,39.3){(c)}
\put(47,39.3){(d)}
\put(-2,60){\begin{turn}{90}{$R_{\delta}^{5/6}\tilde{u}_{\text{max}}$}\end{turn}}
\put(0,20){\begin{turn}{90}{$\delta_{\text{disp}}$}\end{turn}}
\end{overpic}
\caption{
PCS results for various Reynolds numbers. Panels (a) and (c) show the same computational results, for the wave amplitude and the displacement thickness at the peak vortex location, respectively. The computations are performed from the linear critical point $x^*\approx 95$ [cm] using $\Omega \approx 0.1$ 
($\Omega=0.1080,0.1216,$ and $0.1262$ for $Re=15625, 31250, $ and $40000$, respectively). Panels (b) and (d) present results corresponding to panels (a) and (c), respectively, but at a higher frequency $\Omega \approx 0.2$ ($\Omega=0.2079,0.2126,0.2180,$ and $0.2269$ for $Re=31250, 40000, 50000, $ and $60000$, respectively).
}
\label{fig:high_low_different_R_combine}
\end{figure}
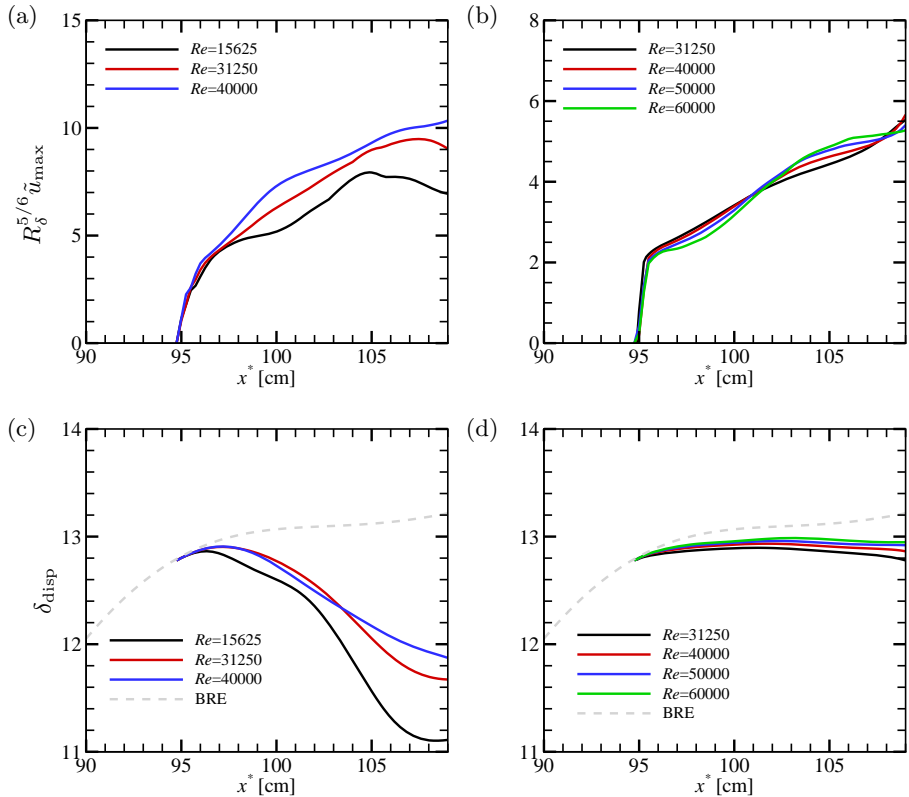

\textcolor{black}{
Finally, we examine the dependence of the computational results on the Reynolds number.
The PCS system retains not only the leading order terms but also higher order terms. Therefore, studying whether the PCS solution attains asymptotic convergence is worthwhile.
The red line in figure 
\ref{fig:high_low_different_R_combine}-(a) shows the same result as the 
$f^* = 170$ [Hz] in figure \ref{fig:fig5}-(a). In addition, we include results for two different $Re$, adjusting $\Omega$ so that the neutral point occurs at approximately $x^*=95$ [cm]. After rescaling $\tilde{u}_{\text{max}}$ by $R_{\delta}^{6/5}$, the two highest $Re$ results lie close each other, at least up to around $x^*=110$ [cm], which is the range of interest in this study.
The scaling factor is consistent to the VWI theory  \citep{hall1991, HALL_SHERWIN_2010}, which shows that the wave amplitude bumps up within the critical layer by a factor of $\epsilon^{-1}$ (i.e. $\epsilon^{-1}\times \epsilon^{1/2}R_{\delta}^{-1}=R_{\delta}^{-5/6}$). 
Figure \ref{fig:high_low_different_R_combine}-(c) shows a similar comparison, but for the displacement thickness at the peaks of the G\"ortler vortices (the same format as figure \ref{fig:fig1}).
Again, a reasonably good level of convergence can be observed.
We performed the same test using a higher frequency, $f^*\approx 280$ [Hz], for which another critical point appears around $x^*=95$ [cm] (open circle in figure \ref{fig:fig2}-(a)).
The results are shown in figures \ref{fig:high_low_different_R_combine}-(b) and (d). This second test shows the asymptotic convergence more clearly, probably because in the first test the neutral curve exhibits relatively complex behaviour at the onset of finite amplitude waves (see figure \ref{fig:fig2}-(a)).
}

\textcolor{black}{\cite{Hall_1988} theoretically showed that the error between the Navier-Stokes and BRE results is very small when $Re$ is large. Therefore, the bottleneck in the asymptotic convergence of the PCS arises from (\ref{waveeqq}), which essentially has the same structure as the equations governing secondary instability. 
The neutral solution of the leading order secondary instability problem was recently solved by \cite{Deguchi_2019} for a model base flow and compared with finite Reynolds number results. Based on this, a Reynolds number of $O(10^4)$, defined using the characteristic length and velocity scales of the base flow, is sufficient to achieve good asymptotic convergence.
In our computation, with a typical  height of the G\"ortler vortices $d^*\approx 1$[cm] (see figure~\ref{fig:fig1}), 
the effective local Reynolds number can be estimated as $d^* U_{\infty}^*/\nu^*\approx O(10^3)$--$O(10^4)$, which supports the observations in figure \ref{fig:high_low_different_R_combine}.}

\textcolor{black}{
Note that in figure \ref{fig:high_low_different_R_combine} the horizontal axis is $x^*=L^*X$, where $L^*=10$ [cm] is used for dimensionalisation. 
Convergence of the wave amplitude/roll-streaks in this coordinate thus implies convergence in the $X$ coordinate.
By contrast, the wavelength is scaled with $\delta^*=Re^{-1/2}L^*$; thus, increasing $Re$ leads to a shorter wavelength in the $x^*$ coordinate. For example, in figure \ref{fig:high_low_different_R_combine}-(b), the typical wavelength is 0.7-1.2 [cm] for $Re=31250$, and  decreases to 0.4-0.7 [cm] for $Re=60000$.
}

\section{\textcolor{black}{Discussion}}\label{sec:4}

\textcolor{black}{This paper presents a theoretical and computational study of the development of finite-amplitude waves on G\"ortler vortices in a boundary layer over a concave wall. Conventionally, the evolution of such waves is explained by tracing the spatial growth of dominant secondary instability modes. In contrast, the PCS determine the wave amplitude by requiring the waves to be self-sustained neutral modes: the Reynolds stresses arising from the interaction of the travelling waves modify rolls and streaks, and the resulting streaks, in turn, render the waves neutral.}

The growth rate $\sigma(X)=\frac{1}{\tilde{u}_{\text{max}}}\frac{d \tilde{u}_{\text{max}}}{dX}$ computed by the slow-scale variation of the wave amplitude 
$\tilde{u}_{\text{max}}$ \textcolor{black}{agrees better with the SB87 experimental results than the prediction from the secondary instability analysis (figure \ref{fig:fig5}).} Moreover, as shown in figure~\ref{fig:fig1}, the PCS method successfully reproduces \textcolor{black}{the displacement thickness observed in the experiments} up to $x^*$=110 [cm], where vortex breakdown \textcolor{black}{to} turbulence is reported.
\textcolor{black}{These results are intriguing, as they show that the PCS approach, which intrinsically incorporates an extension of exact coherent structures, can yield meaningful results for non-parallel flows.}

\textcolor{black}{
While the PCS predictions perform well,  some caution is nevertheless required when interpreting our results. 
In particular, our results are generated using a rather arbitrary initial condition, chosen primarily for simplicity.
G\"ortler vortices can also be generated by perturbations passively advected from upstream in the oncoming uniform flow \citep{WU_ZHAO_LUO_2011,Xu_Zhang_Wu_2017} or coherent structures self-sustained in free-stream \citep{deguchi2014,deguchi_hall_2017,DHD_2017}. Using those more realistic and complex conditions may lead to improved predictions for the BRE part. Also, the PCS approach requires the value of $\Omega$ to be prescribed. In the experiment the frequency of the wave like activity appears to be set an external perturbation, the nature of which is not described in SB87. 
We therefore estimated $\Omega$ from the experiments and induced the waves using a small artificial forcing, as described in Appendix~\ref{sec:AppA}. If desired, this forcing can be replaced by physical mechanism, for example that generated by an oscillatory pressure gradient and a sinusoidal wavy wall as considered in \cite{Luo_Wu_2004}. Consequently, a connection between receptivity analysis in the boundary layer research community and the PCS approach may be explored in future studies.
}

\textcolor{black}{
We also remark that the conventional transition route via the modally unstable modes certainly play a major role in some other roots of boundary layer transitions (see the review papers by \cite{Reed_1996} and \cite{Saric2003} for flat-plate or swept-wing boundary layers). The absence of this behaviour in the SB87 experiment can be explained by two factors: (i) the experimental setup is subjected to relatively large oncoming disturbances, and (ii) in the G\"ortler vortex problem, even very small wave amplitudes can significantly affect the streaks. }

\textcolor{black}{The need to consider factor (i) is apparent from figure~\ref{fig:fig5}-(a), where $\tilde{u}_{\text{max}}$ is not small, being of order $O(10^{-2})$, even upstream of the transition point. 
The external perturbation level can be more easily controlled in DNS, and \cite{Souza_2017} reported a case in which the upstream $\tilde{u}_{\text{max}}$ is $O(10^{-5})$.
Using the parameters from this simulation, we repeated the BRE, PCS, and secondary instability computations, as shown in Appendix~\ref{sec:AppB}. When the wave amplitude is extremely small, the secondary instability provides reasonably good results, whereas the wave amplitude approaches the saturation level, the PCS yields significantly better predictions.}

\textcolor{black}{Factor (ii) can be understood by a feature of VWI theory that even very small wave amplitudes significantly affect the streaks. As is well known, VWI inherently contains the self-sustained process \citep{hamilton1995,waleffe1997}, in which the waves drive the rolls through the Reynolds stress, the rolls modify the streaks via the lift-up mechanism, and the resulting modulation of the streaks then influences the waves. Among these, the lift-up mechanism is the key for amplifying small perturbations, and it can exist only in three-dimensional flows. VWI is also relevant to bypass transition, and it is noteworthy that the scenario described here may be applicable.
}

\textcolor{black}{
The above argument implies that exponentially growing perturbations rapidly violate the small-amplitude assumption underlying linear analysis and thus persist only over short spatiotemporal scales. 
From a physical perspective, the locally neutral solution is therefore the one that tends to survive the longest.
Indeed, a closer look at the analysis of section \ref{sec:2.1} shows that}
local neutrality is an essential requirement for constructing the rational asymptotic theory; without it, the Reynolds stress would depend on the fast variable $\theta$, contradicting the assumption that the roll–streak field $\overline{\mathbf{u}}$ is independent of $\theta$. The consistency with the VWI  is precisely what enables the PCS method to combine the efficiency of the spatial marching approach with the robustness of exact coherent-structure computations.

Finally, we note that the significance of our results also lies in the long-awaited achievement of the first numerical computation of VWI in spatially growing boundary-layer flows, originally formulated by \cite{hall1991}. 
The solution obtained in \S~\ref{sec:3} approximates a time-periodic exact coherent structure, which is very likely unstable.
It is noteworthy that, in the parallel flow research community, such unstable periodic orbits are now recognised as fundamental building blocks for understanding turbulence from the perspective of dynamical systems theory perspective \citep{Kawahara_2012}. 
We expect that a similar approach will, in the future, become increasingly popular for realistic flows with non-parallel effects, where the efficient computational strategy of the PCS will play a key role.

\backsection[Acknowledgements]{
This research was supported by the Australian Research Council Discovery Project DP230102188. }

\backsection[Declaration of Interests]{
The authors report no conflict of interest.
}

\appendix
\section{Generation of finite amplitude waves by external forcing}\label{sec:AppA}

\begin{figure}
\centering
\begin{overpic}[width=0.9 \textwidth]{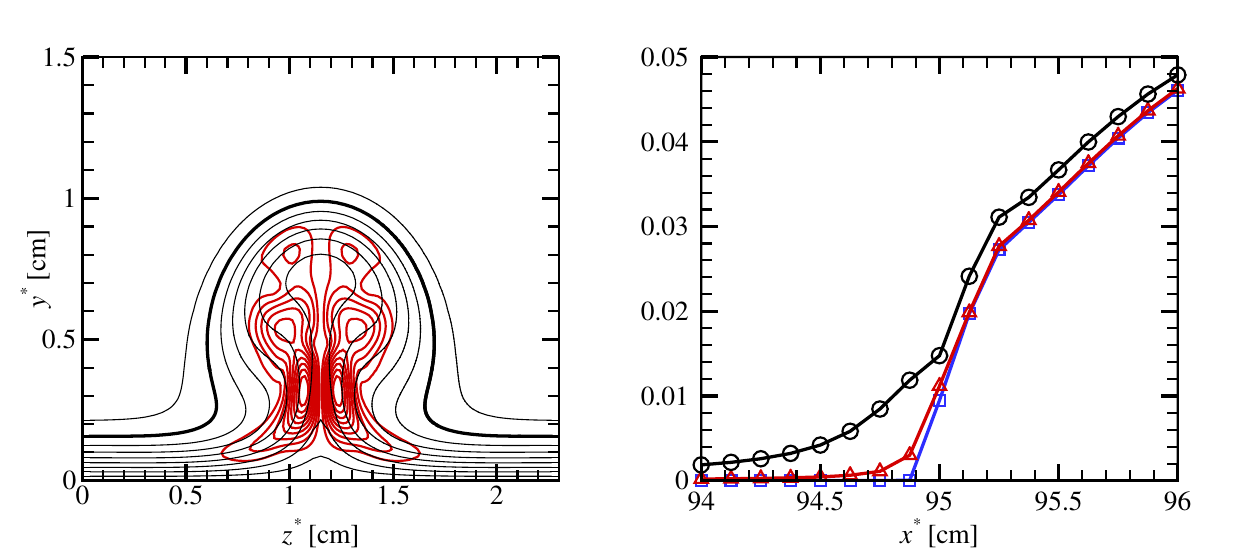}
\put(48,20){\begin{turn}{90}{$\tilde{u}_{\text{max}}$}\end{turn}}
\put(-3,39.3){(a)}
\put(47,39.3){(b)}
\end{overpic}
\caption{
Analysis near the linear critical point of the second odd mode at $f^*=170$ [Hz].
(a) The red lines are contours of $|\hat u|$
associated with the neutral eigenfunction found by the secondary instability analysis. 
The black lines denote the contours of the base flow $\bar{u}$ at the linear critical point (the same format as figure~\ref{fig:fig4_new}). 
(b) The PCS computation near the linear critical point $x^*=94.88$ [cm]. The external forcing term is added to (\ref{waveeq}). 
The blue, red and black lines correspond to $A=2\times 10^{-5}$, $2\times 10^{-4}$, and  $2\times 10^{-3}$, respectively.}
\label{fig:different_forcing_amp}
\end{figure}

\textcolor{black}{
Here we elaborate on the forcing approach for obtaining a finite amplitude wave solution in the PCS system. 
All the results presented here are obtained for  $f^*=170$ [Hz]. For this frequency, the secondary instability analysis identifies a zero growth rate for the second odd mode at $x^* \approx 94.88$ [cm]. 
Figure~\ref{fig:different_forcing_amp}-(a) shows the corresponding neutral eigenfunction, normalised so that $\tilde{u}_{\text{max}}=1$. We denote this normalised function by $[\tilde{\mathbf{u}}_c,\tilde{p}_c]$.
}

\textcolor{black}{
Now let us consider the spatial marching step at the neutral point. In the BRE computation, the Newton-Raphson method yields the updated field $[\overline{\mathbf{u}},\overline{p}]$. 
This is, of course, the solution of the PCS system with no wave component. 
We then substitute $[A\tilde{\mathbf{u}}_c,A\tilde{p}_c]$ with the prescribed finite amplitude $A$; this introduces a small but non-zero residual.
The residual in the wave momentum equation (\ref{waveeq}) is the external forcing term mentioned at the end of section~\ref{sec:2}. That is, the forcing enters (\ref{waveeq}) as an inhomogeneous term, and  is a function of $A$.
}

\textcolor{black}{
Figure~\ref{fig:different_forcing_amp}-(b) shows the spatial marching result near the neutral point, where the forcing is switched on at $x^*=94$ [cm]. As shown by the blue curve even a very small value of $A=2\times 10^{-5}$ allows the finite amplitude wave solution branch to be captured. The frequency of the PCS solution emanated from the neutral point matches that of the applied forcing. When the amplitude is increased to $A=2\times 10^{-4}$ (red curve), the effect of the forcing becomes apparent in the vicinity of the neutral point. 
In dynamical system analysis, the region near a bifurcation point is often most sensitive to inhomogeneous terms. This phenomenon is known as imperfect bifurcation, which we have exploited in our approach.
The black curve corresponds to an even larger amplitude, $A=2\times 10^{-3}$; at this forcing level, the small deviation from the blue curve persists downstream.
}

\textcolor{black}{
We have tested several different transverse structures for the forcing and obtained qualitatively the same results; as long as the amplitude is small, the effect of the forcing does not appear downstream. In section~\ref{sec:3}, we use the result obtained with $A=2\times 10^{-5}$ in figure~\ref{fig:high_low_different_R_combine}-(b) and switch off the forcing at $x^*=95.5$ [cm]. 
}

\section{Comparison with DNS}\label{sec:AppB}

\begin{figure}
\centering
\begin{overpic}[width=0.9 \textwidth]{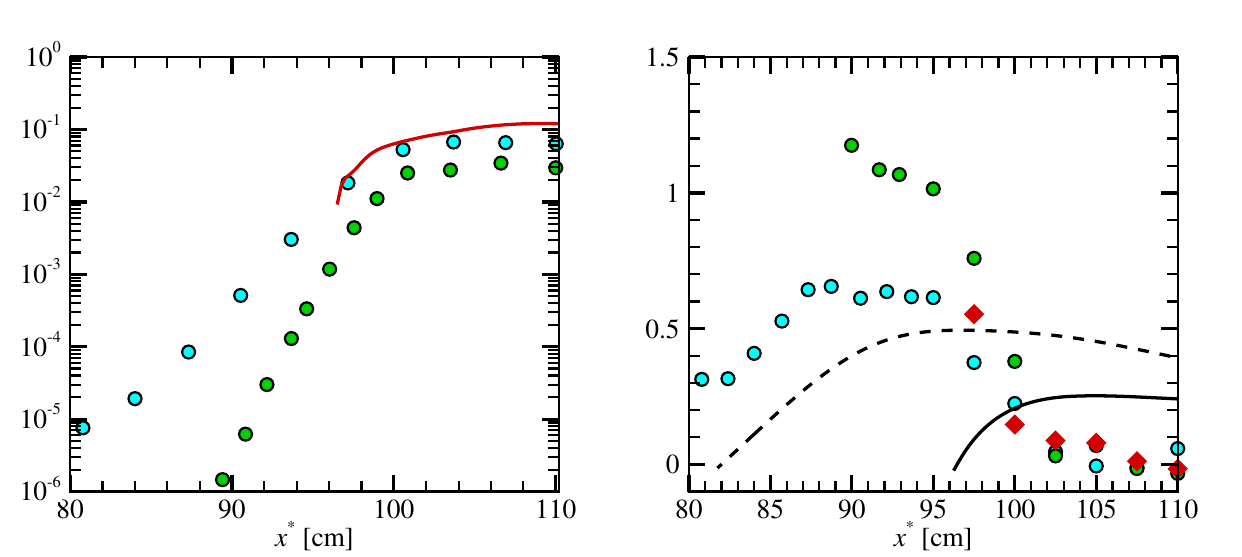}
\put(-2,20){\begin{turn}{90}{$\tilde{u}_{\text{max}}$}\end{turn}}
\put(48,18){\begin{turn}{90}{$\sigma^{*}$ [cm$^{-1}$]}\end{turn}}
\end{overpic}
\caption{Comparison with DNS. 
(a) Amplitude of the wave field. Blue and green circles represent the DNS data reported by \cite{Souza_2017} for 120 and 180 [Hz], respectively. The red line shows the PCS result for $f^*=180$ [Hz]. (b) Growth rate. The circles and diamonds are computed from the DNS and PCS data, respectively. The lines are the secondary instability analysis results. The dashed line is the first odd mode at 120 [Hz]; the solid line is the second odd mode at 180 [Hz].
}
\label{fig:DNS_comparison}
\end{figure}

\textcolor{black}{
Figure \ref{fig:DNS_comparison} presents a comparison similar to that in figure \ref{fig:fig5}, using the DNS of \cite{Souza_2017}. In the DNS,
the G\"ortler vortices are first generated by a steady disturbance generator, and 
an unsteady perturbation is then introduced at $x^* \in [75.05,78.25] $ [cm] via a second disturbance generator.
The second generator is driven by oscillatory blowing and suction on the wall, producing a wide range of frequencies (from 20 to 320 [Hz]). In figure 13 of \cite{Souza_2017}, the downstream evolution of the wave amplitude for each frequency is recorded.
Initially, low frequency modes near 120 [Hz] are excited around $x^*=80$ [cm], followed by the higher frequency modes, such as those around 180 [Hz]. The green and blue circles in figure \ref{fig:DNS_comparison} are representative data from DNS.
}

\textcolor{black}{
\cite{Souza_2017} used the spanwise period $\lambda^*=1.8$ [cm]
and the Reynolds number $Re=33124$, which differ slightly from those used in section \ref{sec:3}. 
Therefore, we produced the PCS results using these parameters as shown by the red line in figure \ref{fig:DNS_comparison}-(a). We used $f^*=180$ [Hz], for which the linear critical point exists around $x^*=96$ [cm]. Since the forcing approach is used (see Appendix~\ref{sec:AppA}), we show only the amplitude levels where the results are unaffected by the forcing. 
The red diamonds in figure \ref{fig:DNS_comparison}-(b) represent the growth rates calculated from PCS and show good agreement with the results from DNS when the wave amplitude is sufficiently large. Note that some discrepancy is expected, because the BRE computation uses rather arbitrary analytic initial condition as mentioned in section ~\ref{sec:2}. Introduction of the steady disturbance generator used by \cite{Souza_2017} would require substantial modifications to the BRE code, which are beyond the scope of this paper.
}

\textcolor{black}{
We also performed secondary instability analyses and confirmed that the results are similar to those shown in figure \ref{fig:fig2}-(a). The first odd mode appears initially, and its growth rate at 120 [Hz] is obtained as the black dashed line in figure \ref{fig:DNS_comparison}-(b). This result appears reasonably close to the DNS results when the wave amplitude remains very small. We also examined the second odd mode at 180 [Hz] (black solid line), but it does not yield particularly meaningful results.
}

\bibliographystyle{jfm}  
\bibliography{Reference}  
\end{document}